\documentclass[12pt]{iopart}
\usepackage{iopams}
\usepackage{graphicx}

\begin{document}

\title{Mechanical entanglement via detuned parametric amplification}

\author{A Szorkovszky$^1$, AA Clerk$^2$, AC Doherty$^3$, and WP Bowen$^1$}
\address{$^1$ Centre for Engineered Quantum Systems, University of Queensland, St Lucia, Australia}
\address{$^2$ Department of Physics, McGill University, Montr\'eal, Canada}
\address{$^3$ Centre for Engineered Quantum Systems, University of Sydney, Sydney, Australia}
\ead{alexs@physics.uq.edu.au}

\begin{abstract}
We propose two schemes to generate entanglement between a pair of mechanical oscillators using parametric amplification. In contrast to existing parametric drive-based protocols, both schemes operate in the steady-state. Using a detuned parametric drive to maintain equilibrium and to couple orthogonal quadratures, our approach can be viewed as a two-mode extension of previous proposals for parametric squeezing. We find that robust steady-state entanglement is possible for matched oscillators with well-controlled coupling. In addition, one of the proposed schemes is robust to differences in the damping rates of the two oscillators.
\end{abstract}

\section{Introduction}

Observing quantum entanglement between massive objects has been a long-standing milestone in exploring the quantum to classical transition\cite{diosi,julsgaard}, constructing hybrid quantum information systems\cite{regal,rabl,braunstein} and sensing forces with ultrahigh precision\cite{mancini}. This goal has prompted interest in the scaling-up of matter-wave interferometers\cite{gneiting,kofler}, in levitating microparticles\cite{chang} and in arrays of mechanical oscillators\cite{xuereb,schmidt}, among other research. Optomechanical systems, in which mechanical oscillators are coupled to optical or microwave fields, are well known as a promising basis for observing macroscopic entanglement in a wide variety of architectures\cite{marquardt,aspelmeyer}.

The advantage of optomechanics lies in the ability to effectively transduce mechanical motion; using backaction evading methods\cite{braginsky}, in principle one can achieve precision beyond the level of the quantum zero-point motion\cite{clerk}. Measurement with sub-zero point precision is only possible for one quadrature of motion, at the expense of degraded sensitivity in the other, due to Heisenberg's uncertainty principle. Such ``quantum squeezing'' of a quadrature, when applied to collective observables of two or more oscillators, yields quantum entanglement between the oscillators. Recently, the theory of optomechanical back-action evasion, which allows measurement-based squeezing, has been expanded to two-mode systems\cite{woolley}, providing a route to mechanical entanglement. In addition, entanglement can be achieved via dispersive\cite{mancini2,paternostro,seok} and dissipative\cite{wang,tan} interactions with cavity fields, including the use of squeezed and entangled fields\cite{zhang,pinard,abdi}. However, while feasible in principle, purely optomechanical entanglement is difficult to achieve in practice due to the requirement of strong and efficient coupling to the optical or microwave field.

In parallel to developments in optomechanics, the fabrication of arrays of electromechanical resonators has developed to an extent that multi-mode coupling can now be precisely controlled\cite{faust,okamoto}.  Theoretical work shows that modulation of these couplings, which reduces fluctuations in certain collective quadratures of motion, is also sufficient for the creation of entanglement\cite{tian,galve}, thus providing a simple and accessible alternative to purely measurement-based schemes. In this previous work, appreciable entanglement could not be sustained for steady-state operation, as is also the case for resonant parametric squeezing of a single oscillator\cite{blencowe}.

We have recently shown that with the aid of weak continuous measurement, detuning a parametric drive from resonance allows strong steady-state squeezing of an oscillator\cite{njp}. Here, we show that the same principle can be applied generally to coupled oscillators in at least two feasible scenarios, allowing strong two-mode entanglement. The first scheme involves a modulation of the coupling between the oscillators, while the second combines a constant linear coupling with single-mode parametric drives. Since only weak continuous measurement is required, the oscillators can be monitored individually without spurious back-action noise, thereby avoiding the need for restrictive measurement setups that couple only to the collective variables of interest. Additionally, the inbuilt parametric tunability of the oscillators in our scheme relaxes the engineering requirements for the physical device. Finally, we show that the entanglement generated by our scheme can be achieved with realistic experimental parameters, and compare this to back-action evading methods\cite{woolley}, making use of Duan's inseparability criterion\cite{duan}.

\section{Model One: Modulated Coupling}

\begin{figure}[bt]
\centering
\includegraphics[width=12cm]{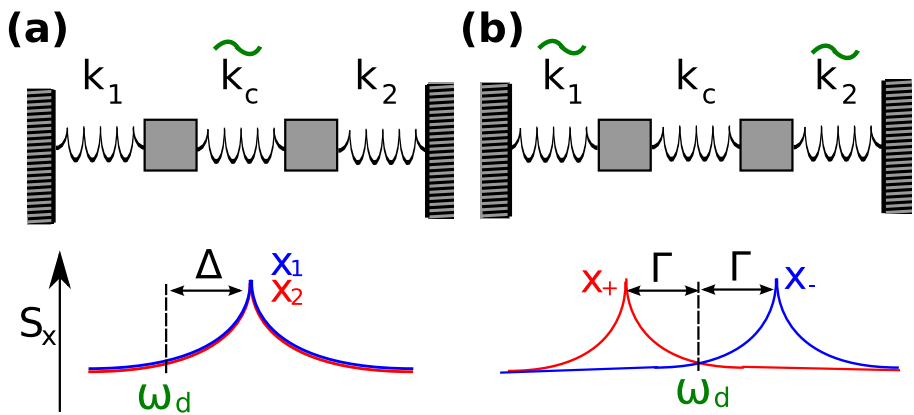}
\caption{\label{schematic}Two approaches to achieving entanglement of mechanical oscillators. Each approach is idealised as a coupled mass-on-spring system, with time-dependent modulations at frequency $2\omega_d$ of (a) the intermodal coupling $k_c$ or (b) the spring constants $k_1$ and $k_2$. Below are plots of the noise spectra of the mechanical modes in the absence of parametric driving. In (a) the mechanical modes of the two masses are degenerate while in (b) there is normal-mode splitting. In addition, the relation between $\omega_d$ and the detuning parameters $\Delta$ and $\Gamma$ are indicated.}
\end{figure}

Here, we consider a simple detuning of a previously proposed mechanical two-mode squeezing scheme\cite{tian} involving a modulated position-position coupling. Consider two oscillators with identical resonance frequency $\omega_m$, which have a controlled time-dependent coupling between them as shown in Figure \ref{schematic}(a). If the coupling is sinusoidally modulated about zero at a frequency $2\omega_d$, the Hamiltonian can be written
\begin{equation}
\hat H = \frac{1}{2m}(\hat p_1^2 + \hat p_2^2) + \frac{m\omega_m^2}{2}(\hat x_1^2 + \hat x_2^2) + g \hat x_1 \hat x_2 \cos(2\omega_d t) \; ,
\end{equation}
where the half-modulation frequency $\omega_d = \omega_m-\Delta$. We focus on a modulation frequency near $2\omega_m$, such that $\Delta \ll \omega_m$. Going into the rotating frame at $\omega_m-\Delta$ after making the usual tranformation to annihilation operators $a$ and $b$ for the two oscillators yields
\begin{eqnarray}
\tilde H &=& \hbar\Delta(\hat a^\dag \hat a + \hat b^\dag \hat b + 1) + \hbar\chi (\hat a \hat b + \hat a^\dag \hat b^\dag) \\
 &=& \frac{\hbar\Delta}{2}(\hat X_1^2 \!+\! \hat Y_1^2\! + \!\hat X_2^2\! +\! \hat Y_2^2) + \hbar\chi (\hat X_1 \hat X_2\! -\! \hat Y_1 \hat Y_2) \; , \nonumber
\end{eqnarray}
where the single-mode quadrature operators are defined as $\hat X_1=(\hat a+\hat a^\dag)/\sqrt{2}$, $\hat Y_1=-i(\hat a-\hat a^\dag)/\sqrt{2}$ and similarly for the other oscillator. We can see that the terms proportional to the parameter $\chi=g / 2m\omega_m$ constitute a two-mode squeezing operation, well-known in quantum optics\cite{heidmann}. In the resonant case $\Delta=0$, this two-mode squeezing results in amplification of two collective quadratures of motion and squeezing of the other two. When the rate of this process exceeds the damping rate of the system (i.e. $\chi>\gamma$), the amplification causes exponential growth of the mechanical oscillations, leading to deleterious mechanical and measurement nonlinearities. Tian et al.\ overcome this by following the squeezing process with a second modulation to swap the fluctuations between quadratures\cite{tian}. Here, we instead consider the steady-state behaviour in the more general case where $\Delta\neq 0$.

We define the quadratures as those of the two natural collective modes $x_+ = (x_1+x_2)/\sqrt{2}$ and $x_- = (x_1 - x_2)/\sqrt{2}$, namely
\numparts
\begin{eqnarray}
\hat X_+ &=& (\hat X_1+\hat X_2)/\sqrt{2} \\
\hat X_- &=& (\hat X_1-\hat X_2)/\sqrt{2} \\
\hat Y_+ &=& (\hat Y_1+\hat Y_2)/\sqrt{2} \\
\hat Y_- &=& (\hat Y_1-\hat Y_2)/\sqrt{2}  \; .
\end{eqnarray}
\endnumparts
The two non-zero commutators between these operators are
\begin{equation}
[\hat X_+,\hat Y_+] = [\hat X_-,\hat Y_-] = i \; .
\end{equation}
In these collective variables, the Hamiltonian can be factorised as
\begin{equation}
\tilde H = \frac{\hbar(\Delta+\chi)}{2}(\hat X_+^2 + \hat Y_-^2) + \frac{\hbar(\Delta-\chi)}{2}(\hat X_-^2 + \hat Y_+^2) \; .
\end{equation}

With equal dissipation for both oscillators at rate $\gamma$, the equations of motion are
\begin{equation}
\fl\left[\! \begin{array}{c}
\mathrm{d}\hat X_+ \\
\mathrm{d}\hat X_- \\
\mathrm{d}\hat Y_+ \\
\mathrm{d}\hat Y_- \end{array}\! \right] \!=\! 
\left[\begin{array}{cccc}
        -\gamma & 0 & -\chi\!+\!\Delta & 0\\
        0 & -\gamma & 0 & \chi\!+\!\Delta\\
        -\chi\!-\!\Delta & 0 & -\gamma & 0\\
        0 & \chi\!-\!\Delta & 0 & -\gamma
       \end{array}\right] \left[\! \begin{array}{c}
\hat X_+ \\
\hat X_- \\
\hat Y_+ \\
\hat Y_- \end{array}\! \right]\mathrm{d}t +\!\sqrt{2\gamma}\left[\! \begin{array}{c}
\mathrm{d}\hat X_{+\mathrm{in}}(t) \\
\mathrm{d}\hat X_{-\mathrm{in}}(t) \\
\mathrm{d}\hat Y_{+\mathrm{in}}(t) \\
\mathrm{d}\hat Y_{-\mathrm{in}}(t) \end{array}\! \right] \; .
\end{equation}

It is easy to see that the four collective quadratures can be sorted into their non-commuting pairs, with the two pairs independent of each other
\begin{equation} \label{meanvalue}
\left[\! \begin{array}{c}
\!\mathrm{d}\hat X_\pm\! \\
\!\mathrm{d}\hat Y_\pm\! 
 \end{array} \!\right]\!=\!\left[\begin{array}{cc} 
        \!-\gamma & \Delta\!\mp\!\chi\! \\
        \!-\Delta\!\mp\!\chi & -\gamma\!
       \end{array}\right] \!
\left[\! \begin{array}{c}
\!\hat X_\pm\! \\
\!\hat Y_\pm\! \end{array} \!\right]\!\mathrm{d}t  +\!\sqrt{2\gamma}\!\left[\! \begin{array}{c}
\!\mathrm{d} \hat X_{\pm \mathrm{in}}(t)\! \\
\!\mathrm{d} \hat Y_{\pm \mathrm{in}}(t)\! \end{array} \!\right]
\end{equation}

It is important to note that the independence of each pair of quadratures requires the two oscillators $\omega_m$ to have identical resonance frequencies and damping rates $\gamma$. While the resonance frequencies in this scheme can be made equal by using an optical spring\cite{marquardt} or capacitive tuning\cite{rugar,paramp1} on individual oscillators, the individual damping rates are more difficult to engineer. Unequal frequencies or damping rates would be expected to degrade entanglement, the analysis of which would require the solution to the full 10-element covariance matrix in the case of modulated coupling. The constant coupling scheme below is more naturally robust to these experimental imperfections, as will be further discussed in Section \ref{asym}.

The system we have described remains below threshold and therefore convergent as long as $\chi < \chi_\mathrm{th}$ where $\chi_\mathrm{th}=\sqrt{\gamma^2 + \Delta^2}$. That is, $\chi$ can be made much larger than $\gamma$ as long as the absolute detuning follows suit.

To see the utility of a large detuning, consider the simple case $\Delta = \pm\chi$. For example, setting $\Delta=\chi$ gives
\begin{equation}
\tilde H = \hbar\chi (\hat X_+^2 + \hat Y_-^2) \; .
\end{equation}
This Hamiltonian is similar to a quantum non-demolition (QND) measurement in that in the absence of damping, $\hat X_+$ and $\hat Y_-$ are constants of the motion. The significance of this scenario is discussed in Ref.\ \cite{qnd} for a single continuously measured oscillator, where it is shown to correspond to backaction evading measurement. As can be verified from Eq.\ (\ref{meanvalue}), the quasi-QND variables $\hat X_+$ and $\hat Y_-$ are now only influenced by dissipation. However, a time-dependent signal --- such as thermal noise --- in $\hat X_+$ or $\hat Y_-$ will appear in the subsequent evolution of non-QND observables $\hat Y_+$ and $\hat X_-$, respectively. For values of $\chi$ much greater than $\gamma$, these signals will appear strongly amplified. Weak measurements of $\hat Y_+$ and  $\hat X_-$ then provide enhanced effective measurements of the quasi-QND observables $\hat X_+$ and $\hat Y_-$. The result is that one can strongly condition the latter quadratures without the backaction of a strong measurement. The enhanced collective measurement described above also occurs in the general case $|\Delta|\neq\chi$, albeit without the simplified QND dynamics, and is most pronounced at the slightly lower detuning where $\chi \approx \chi_\mathrm{th}$.

\section{Collective-mode Measurement and Entanglement}

\label{collmeas}

For ideal continuous variable entanglement, it is necessary for two collective quadratures of the two oscillators to be localised to below the zero-point motion. In principle, two commuting collective quadratures can be measured without backaction, implying that this two-mode squeezing can be achieved by a strong measurement. However, achieving such an ideal non-local measurement is difficult without complex measurement techniques. For example, the variable $\hat x_1-\hat x_2$ can be measured by using the two oscillators as the end-mirrors of an optomechanical cavity. However, this configuration only yields information about the quadratures $\hat X_-$ and $\hat Y_-$, which do not commute and hence cannot be squeezed. The measurement of only two commuting collective observables is the task of two-mode backaction evasion\cite{woolley}, which will be discussed later.

Detuned parametric driving provides an alternative solution to this problem by making sub-zero point collective fluctuations accessible to weak measurement. This is possible due to correlations between the squeezed and amplified collective quadratures, with the specific dynamics shown in the previous section. With only weak measurement, the position of each oscillator can be independently and continuously monitored without significant backaction, eliminating the need for specialised measurement techniques. Instead, collective quadratures can be localised to below the zero point level by conditioning on the classical measurement records. In other words, while a squeezed quadrature (for example $X_+$) has its own measurement record dominated by measurement noise, an accurate estimate of $X_+$ can be filtered from the fluctuations of its amplified pair $Y_+$, which can be well transduced. In this sense, detuned parametric amplification with weak measurement conditioning is an alternative to two-mode backaction evading measurement.

\section{Conditional Variances}

The conditional variance quantifies the error in the optimal estimate of an observable when past measurements are taken into account. This is equivalent to the confinement achievable when using this optimal estimate for negative feedback, although localization due to the measurement itself is sufficient to confirm entanglement. Quantum mechanically, the conditional variance can be obtained using a stochastic master equation\cite{jacobs,njp} that models the effects of measurement as well as thermal noise. Let us assume the two oscillator positions $x_1$ and $x_2$ are measured independently. All four collective quadratures will then experience the same rate of back-action and conditioning, quantified by the parameter $\mu$. An observable $\hat A$ will evolve as
\begin{eqnarray}\label{master}
\fl\mathrm{d}\langle\hat A\rangle &=& -\frac{i}{\hbar}\langle[\hat A,\tilde H]\rangle\,\mathrm{d}t + [2\gamma N+\mu]\langle\mathcal{D}[\hat a^\dag + \hat b^\dag]\hat A\rangle\,\mathrm{d}t +  [2\gamma(N+1)+\mu]\langle\mathcal{D}[\hat a + \hat b]\hat A\rangle\,\mathrm{d}t \nonumber \\
\fl&& + \sqrt{\eta \mu}(\langle\mathcal{H}[\hat X_+]\hat A\rangle\,\mathrm{d}W_1 \!+\! \langle\mathcal{H}[\hat Y_+]\hat A\rangle\,\mathrm{d}W_2 \!+\! \langle\mathcal{H}[\hat X_-]\hat A\rangle\,\mathrm{d}W_3 \!+\! \langle\mathcal{H}[\hat Y_-]\hat A\rangle\,\mathrm{d}W_4) \, ,
\end{eqnarray}
where $N$ is the mean bath phonon number and $\eta$ is the quantum efficiency. The superoperator
\begin{equation}
\mathcal{D}[\hat a]\hat A = \hat a^\dag\hat A\hat a - \frac{1}{2}(\hat a^\dag \hat a\hat A + \hat A\hat a^\dag \hat a)
\end{equation}
describes the thermal diffusion and back-action. The other superoperator is of the form
\begin{equation}
\mathcal{H}[\hat a]\hat A = \hat a\hat A + \hat A\hat a^\dag - \langle \hat a+\hat a^\dag\rangle\langle\hat A\rangle \; ,
\end{equation}
which updates the conditional values of the observables based on the residual noise processes $\mathrm{d}W_n$. The evolution of the conditional variances is obtained by inserting linear and quadratic observables into the master equation\cite{njp}.

In the steady state, the only non-zero covariances are those between $\hat X_+$ and $\hat Y_+$ and between $\hat X_-$ and $\hat Y_-$. This leaves two independent sets of three equations, written in collated form as
\begin{eqnarray}
\fl\frac{\mathrm{d}V_{X\pm}}{\mathrm{d}t} & \!= & -2\gamma V_{X\pm} - 2(\Delta\!\mp\!\chi)C_{\pm} + 2\gamma V_0 - \!4\eta\mu(V_{X\pm}^2\!+\!C_{\pm}^2) \nonumber\\
\fl\frac{\mathrm{d}V_{Y\pm}}{\mathrm{d}t} & \!= & -2\gamma V_{Y\pm} + 2(\Delta\!\pm\!\chi)C_{\pm} + 2\gamma V_0 - \!4\eta\mu(V_{Y\pm}^2\!+\!C_{\pm}^2) \nonumber\\  
\fl\frac{\mathrm{d}C_{\pm}}{\mathrm{d}t} & \!= & -2\gamma C_{\pm} \pm \Delta(V_{X\pm}\!-\!V_{Y\pm}) \pm \chi(V_{X\pm}\!+\!V_{Y\pm}) - 4\eta\mu C_{\pm}(V_{X\pm}\!+\!V_{Y\pm}) \; , 
\end{eqnarray}
where
\begin{equation}
C_{\pm} = \frac{1}{2}\langle \hat X_\pm \hat Y_\pm + \hat Y_\pm \hat X_\pm \rangle \; , 
\end{equation}
and
\begin{equation}
V_0 = N + \frac{1}{2} + \frac{\mu}{2\gamma} \; .
\end{equation}
Remarkably, each of these sets is identical to the three variance equations in the one-mode detuned parametric amplification theory, with steady-state solutions already derived\cite{njp}. This means that the amount of conditional two-mode squeezing generated by a coupling rate modulation of $\chi$ is the same as the single mode squeezing available using a spring constant modulation of $\chi$. The maximum squeezing appears at some angle $\alpha$ in the $X_+,Y_+$ plane, and at 90 degrees to this angle in the $X_-,Y_-$ plane. We will denote the variances of these optimal quadratures as $V_{\alpha+}$ and $V_{\alpha-}$, respectively.

The necessary and sufficient condition for entanglement of bipartite Gaussian states has been derived by Duan\cite{duan} and Simon\cite{simon}. Here, in the spirit of the product criterion for the EPR paradox proposed by Reid\cite{reid}, which itself is a sufficient but not necessary condition for entanglement, we quantify the entanglement using a product form for separability\cite{bowen}
\begin{equation}
S = 2\sqrt{V_{\alpha+} V_{\alpha-}}\; ,
\end{equation}
For states that are symmetric between the two oscillators, this quantity is directly related to the log-negativity by $E_N = -\ln(S)$ for $S<1$. In general, a separability below unity as defined above is a sufficient condition for entanglement. This requires the geometric mean of the quadrature variances $V_{\alpha+}$ and $V_{\alpha_-}$ to be below the level of the zero-point motion. Inserting the single-mode solution for $V_\alpha$ from Ref.\ \cite{njp}, we find
\begin{equation}
S = \frac{\sqrt{(\gamma\!+\!\chi\sin(2\alpha))^2\!+\!4\gamma^2 \mathrm{SNR}}\!-\!\gamma\!-\!\chi \sin(2\alpha)}{2\eta\mu} \; ,
\end{equation}
where the signal to noise ratio
\begin{equation} \label{snr}
\mathrm{SNR} = 2\eta\mu V_0/\gamma = 2\eta\mu(N+1/2)/\gamma + \frac{\eta\mu^2}{\gamma^2} \; ,
\end{equation}
quantifies the ratio of the thermal and backaction induced motion of the oscillator to the measurement noise, and the squeezing angle $\alpha$ satisfies
\begin{equation} \label{alphasol}
\fl\cos2\alpha_1\!=\! \frac{\Delta}{\chi_{th}}\!\left(\!\frac{\chi_{th}^2\!+\!\chi^2\!+\!4\gamma^2\mathrm{SNR}\!-\!\sqrt{(\chi_{th}^2\!-\!\chi^2)^2\!+\!8(\chi_{th}^2\!+\!\chi^2)\gamma^2\mathrm{SNR}\!+\!16\gamma^4\mathrm{SNR}^2}}{2\chi^2}\right)^{\frac{1}{2}}
\end{equation}

\begin{figure}[tb]
\centering
\includegraphics[width=10cm]{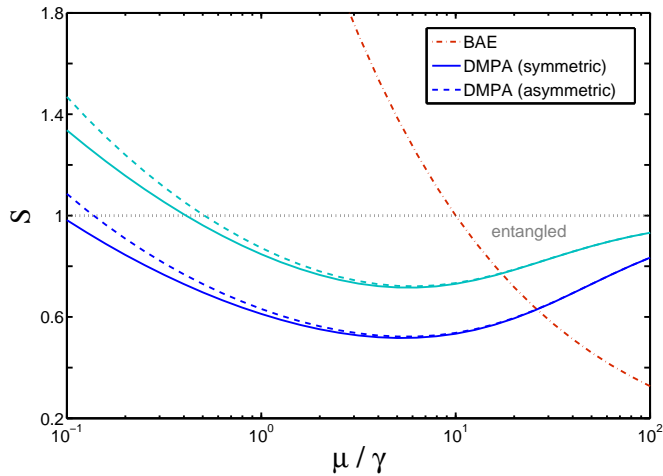}
\caption{\label{duan} Separability $S$ as a function of measurement strength in the case of symmetric damping rates (solid lines); and using the constant-coupling method with a damping asymmetry $\gamma_D = 0.5 \gamma_S$ (dashed lines). Light and dark curves indicate normalised drive strengths $\chi/\gamma$ of 25 and 50, respectively, while the mean phonon occupation $N$ is 5 and efficiency $\eta$ is unity. The dot-dashed curve represents the minimum separability using two-mode backaction evasion.}
\end{figure}

Similar to our previous results for single-mode squeezing\cite{njp}, entanglement is easily achievable for a low mean phonon occupation $N$, detuning situated near threshold ($\chi_{th}\approx \chi$) and with a moderate measurement strength, as shown by Figure \ref{duan}. In the strong measurement regime, backaction causes the $\mu^2$ term in (\ref{snr}) to dominate and the entanglement to disappear as expected. With a moderate measurement strength, the parametric drive boosts the effective measurement into this regime without adding backaction. When $\chi\gg\gamma$, the optimal conditioning occurs near $\mu\approx\gamma(N+1/2)$, with the separability $S$ scaling as $\sqrt{\gamma/\chi}$. In contrast, backaction evasion can only produce entanglement in the strong and efficient measurement regime $\eta\mu\gg\gamma$, as shown in Figure \ref{duan}. 

By combining measurement conditioning and unitary coupling, this scheme hybridises two approaches to entanglement generation. In contrast to purely parametric-based proposals such as Ref.\ \cite{tian}, the entanglement can be made arbitrarily strong with the system remaining in the steady-state. Unlike purely measurement-based proposals, independent weak continuous measurements of the two oscillators are sufficient to generate entanglement between them. This can be confirmed by reconstructing the conditional covariance matrix from the measurement record. Entanglement can also be independently verified via direct tomography of the mechanical states, using strong projective measurements on the individual oscillators\cite{vanner} and analysis of correlations. Even with this verification step, at no stage does a collective mechanical mode need to be measured directly.

\section{Model Two: Constant Coupling}

An equivalent scenario, resulting in the same variance equations, can arise from constant linear coupling between two oscillators, with degenerate parametric drives applied individually to each oscillator on resonance ($\Delta=0$). In this case, as illustrated by Figure \ref{schematic}(b), effective detunings for the collective variables are provided by the normal-mode splitting, which is equal to twice the coupling rate $\Gamma$. The Hamiltonian is given by
\begin{equation}
\fl\hat H = \frac{1}{2m}(\hat p_1^2 + \hat p_2^2) + \frac{\Gamma}{m\omega_m} \hat x_1 \hat x_2 +\frac{m\omega_m}{2}[\hat x_1^2 (\omega_m \!+\! 2\chi \sin 2\omega_m t) \!+\! \hat x_2^2 (\omega_m \!+\! 2\chi \sin 2\omega_m t)] \nonumber
\end{equation}
Going into a rotating frame at $\omega_m$, assuming $\Gamma \ll \omega_m$
\begin{eqnarray}
\tilde H &=& \hbar\Gamma(\hat a^\dag \hat b + \hat a \hat b^\dag) - \frac{i\hbar\chi}{2}(\hat a^2-\hat a^{\dag 2} + \hat b^2 - \hat b^{\dag 2}) \\
 &=& \hbar\Gamma(\hat X_1 \hat X_2 \!+\! \hat Y_1 \hat Y_2) \!+\! \frac{\hbar\chi}{2}(\hat X_1 \hat Y_1 \!+\! \hat Y_1 \hat X_1 \!+\! \hat X_2 \hat Y_2 \!+\! \hat Y_2 \hat X_2) \nonumber
\end{eqnarray}
which factorises as
\begin{equation}
\tilde H = \frac{\hbar(\chi\!+\!\Gamma)}{2}(\hat U_1 \hat V_2 \! + \! \hat V_2 \hat U_1)\! + \!\frac{\hbar(\chi\!-\!\Gamma)}{2}(\hat U_2 \hat V_1 \!+ \!\hat V_1 \hat U_2) \; ,
\end{equation}
where the new collective quadratures are defined as
\numparts
\begin{eqnarray}
\hat U_1 &=& (\hat X_1+\hat Y_2)/\sqrt{2} \\
\hat U_2 &=& (\hat X_1-\hat Y_2)/\sqrt{2} \\
\hat V_1 &=& (\hat Y_1-\hat X_2)/\sqrt{2} \\
\hat V_2 &=& (\hat Y_1+\hat X_2)/\sqrt{2} \; .
\end{eqnarray}
\endnumparts
As in the first model, we consider the independent monitoring of both oscillators, allowing these observables to be constructed trivially via lock-in techniques. It should be noted that the times at which the $\hat X$ and $\hat Y$ quadratures reflect the true position differ by a quarter of an oscillator cycle. However, in the case of high-Q oscillators such that $Q\gg N$, a quarter-cycle is insufficient time for thermal perturbations to influence the oscillators and therefore can be presumed to have little practical effect on entanglement.

The two non-zero commutators between the four new operators are
\begin{equation}
[\hat U_1,\hat V_1] = [\hat U_2,\hat V_2] = i \; ,
\end{equation}
so that with damping, the mean evolution of the four quadratures is given by
\begin{equation}\label{constevol}
\fl\left[ \!\begin{array}{c}
\mathrm{d}\hat U_1 \\
\mathrm{d}\hat U_2 \\
\mathrm{d}\hat V_1 \\
\mathrm{d}\hat V_2 \end{array}\! \right] = \left[\!\begin{array}{cccc}
        -\gamma & \chi\!-\!\Gamma & 0 & 0\\
        \chi\!+\!\Gamma & -\gamma & 0 & 0\\
        0 & 0 & -\gamma & -\chi\!-\!\Gamma\\
        0 & 0 & -\chi\!+\!\Gamma & -\gamma
       \end{array}\!\right]  \left[\! \begin{array}{c}
\hat U_1 \\
\hat U_2 \\
\hat V_1 \\
\hat V_2 \end{array}\! \right]\mathrm{d}t + \!\sqrt{2\gamma}\left[ \!\begin{array}{c}
\mathrm{d}\hat U_{1\mathrm{in}}(t) \\
\mathrm{d}\hat U_{2\mathrm{in}}(t) \\
\mathrm{d}\hat V_{1\mathrm{in}}(t) \\
\mathrm{d}\hat V_{2\mathrm{in}}(t) \end{array}\!\right] \; .
\end{equation}

This immediately resembles Eq.\ \ref{meanvalue} for the case of two-mode squeezing via modulated coupling, but with the detuning $\Delta$ replaced by half of the normal-mode splitting $\Gamma$. This method is analogous to the generation of two-mode squeezing of light by coupling two single squeezed modes on a beam-splitter\cite{bowen2}. Again choosing the simplified case $\chi=\Gamma$, we find that $\hat V_1$ is a proxy observable for $\hat V_2$ and $\hat U_2$ is a proxy observable for $\hat U_1$. The variance equations are likewise identical to those in the modulated coupling case. Notably, unlike the previous method, in this case the dynamical coupling due to the parametric drive is between commuting pairs of observables, that is $[U_1,U_2]=[V_1,V_2]=0$. Therefore, these pairs of observables both qualify as quantum mechanics-free subsystems, a topic of recent interest\cite{tsang}.

\section{Comparison with Measurement-Based Scheme}

Creating two-mode entanglement using parametric amplification and weak measurement avoids the difficult problem of achieving a strong measurement of the commuting quadratures $\hat X_+$ and $\hat Y_-$ without also measuring the orthogonal quadratures $\hat X_-$ and $\hat Y_+$. As described in section \ref{collmeas}, any measurement of the latter quadratures would introduce backaction to (and thus prevent squeezing of) the former. Methods to overcome this backaction using measurement alone generally involve a time-dependent modulation of coupling to the transducer. A recent proposal for this kind of cavity optomechanics-based backaction evasion\cite{woolley} involves using oscillators of differing frequency, such that the quadrature $\hat Y_-$ is measured via dynamic coupling with the single transduced collective quadrature $\hat X_+$. Although the analysis in that work assumes the good-cavity limit, ideal for one-mode backaction evasion\cite{clerk}, spurious heating still arises when the dynamic coupling is too slow. That is, the frequency difference $\Omega = (\omega_b-\omega_a)/2$ must be much larger than the mechanical decay rate for both quadratures to be measured efficiently.

The methods we have outlined here, by contrast, allow strong transduction of two commuting collective quadratures below the level of the zero-point motion, with no extra spurious heating arising from the degeneracy of the oscillators. Instead, the backaction heating is a decreasing function of the parametric drive strength. While in principle $\Omega/\gamma$ can be made very large in the measurement-based scheme by using different sized oscillators, in practice this would lead to asymmetries in the damping and measurement rates and hence further experimental difficulties. In allowing the use of similar oscillators, a parametric amplification scheme for mechanical entanglement offers a significant experimental advantage. In addition, while asymmetry in the measurement coupling must be compensated for in two-mode backaction evasion\cite{woolley}, both of the above methods can be used in the weak measurement regime where unequal backaction noise is less critical. Furthermore, the above methods can be used outside the good-cavity regime of optomechanics, whereas this regime is necessary for two-mode backaction evasion.

\section{Damping Asymmetry}
\label{asym}

To this point the two methods proposed here differ only in experimental implementation. The constant coupling method has the advantage that the ability to tune the individual resonance frequencies is already assumed, and can be achieved by a constant offset of the parametric drive. A more important distinction of the constant coupling method, however, is that it is also robust to unequal damping rates for the two oscillators, a common experimental scenario that cannot otherwise be easily corrected. While the two-mode squeezing method requires equal damping rates $\gamma_1=\gamma_2=\gamma$ to keep the two pairs of quadratures independent of each other, this is not required for the constant coupling method. Instead, the damping asymmetry modifies only the effective parametric drive rate. This can be shown by the more general form of Eq. (\ref{constevol})
\begin{equation}
\fl\left[ \!\begin{array}{c}
\mathrm{d}\hat U_1 \\
\mathrm{d}\hat U_2 \\
\mathrm{d}\hat V_1 \\
\mathrm{d}\hat V_2 \end{array}\! \right] = \left[\!\begin{array}{cccc}
        -\gamma & \chi_1\!-\!\Gamma & 0 & 0\\
        \chi_1\!+\!\Gamma & -\gamma & 0 & 0\\
        0 & 0 & -\gamma & -\chi_2\!-\!\Gamma\\
        0 & 0 & -\chi_2\!+\!\Gamma & -\gamma
       \end{array}\right] \left[\! \begin{array}{c}
\hat U_1 \\
\hat U_2 \\
\hat V_1 \\
\hat V_2 \end{array}\! \right]\mathrm{d}t + \!\sqrt{2\gamma}\left[ \!\begin{array}{c}
\mathrm{d}\hat U_{1\mathrm{in}}(t) \\
\mathrm{d}\hat U_{2\mathrm{in}}(t) \\
\mathrm{d}\hat V_{1\mathrm{in}}(t) \\
\mathrm{d}\hat V_{2\mathrm{in}}(t) \end{array}\!\right] \; ,
\end{equation}
where
\begin{eqnarray}
\chi_1 &=& \chi - (\gamma_1 - \gamma_2)/2 \\
\chi_2 &=& \chi + (\gamma_1 - \gamma_2)/2 \\
\gamma &=& (\gamma_1 + \gamma_2)/2 \; .
\end{eqnarray}
Therefore, an increased effective rate $\chi_2$ drives the squeezing in one pair of quadratures, while in the other pair the rate is decreased to $\chi_1$. At first glance this would appear to have little effect on the separability, since to first order the loss of squeezing in $\hat U_1$ would be made up for by increased squeezing in $\hat V_2$. However, the fact that all quadratures share an effective detuning rate $\Gamma$ implies that when $\hat V_1,\hat V_2$ are at the instability threshold (i.e.\ optimally squeezed), the pair $\hat U_1,\hat U_2$ are bound to be further away from it due to a weaker effective parametric drive. For a strong parametric drive, this asymmetry has a modest effect, as shown for a 50\% difference in damping rates in Figure \ref{duan}.

\section{Experimental Outlook and Conclusion}

We have provided two routes to robust mechanical entanglement, neither of which rely on being in the deeply backaction-dominated regime, being in the optomechanical good-cavity limit, or on temporally modulated measurement coupling. Instead, by using simple parametric processes to create amplified proxy observables, weak or inefficient measurement is sufficient to strongly condition collective quadratures, thus avoiding the need for collective backaction evading measurements. Methods to continually alternate the sign of intermodal coupling between oscillators, as required by the modulated coupling scheme, have been outlined for electronic resonators\cite{tian}, but are difficult to extend to multiple mechanical modes. In contrast, the key technique required for the constant-coupling method (single-mode parametric amplification) is well developed in micromechanical and nanomechanical systems\cite{rugar,paramp1,paramp3}. In such devices, intrinsic intermodal coupling through the substrate is substantial, fulfilling the second requirement of this scheme. An example of such a device, recently demonstrated by Okamoto et al.\cite{okamoto}, has an intrinsic coupling rate $\Gamma$ exceeding the damping rate by a factor of around 400, while the piezoelectric strain is sufficient to parametrically drive well above threshold by at least a factor of 100. For such devices to achieve quantum entanglement using the constant coupling method would require significant improvements in measurement sensitivity, as well as increased mode frequency to reduce the mean phonon occupation. The recent experimental work by Bochmann et al.\cite{bochmann}, in which a high-frequency mechanical resonator has both piezoelectric and optomechanical coupling, appears to be very promising in this respect. By extending such technology to multiple oscillators, steady-state entanglement of massive objects could be well within reach.

\ack
This research was funded by the Australian Research Council Centre of Excellence CE110001013 and Discovery Project DP0987146. AAC acknowledges support from the DARPA ORCHID program under a grant from the AFOSR. WPB acknowledges DARPA via a grant through the ARO. We would also like to thank the referees for helpful comments.

\end{document}